\documentclass[journal]{IEEEtran}

\usepackage{relsize,balance,lipsum,bbm,enumerate,times,comment,color,graphicx,setspace,bbm,mathdots,mathrsfs,amssymb,latexsym,amsfonts,amsmath,cite,stmaryrd,caption,pgf,tikz,accents,mathtools,tabu,enumitem,hhline,array,epstopdf}
\usepackage[ruled,vlined]{algorithm2e}
\usepackage{multirow}
\usepackage[utf8]{inputenc}
\usepackage{dirtytalk}
\graphicspath{{./images/}}
\usepackage[font=small,skip=0pt]{caption}

\title{A Novel Event-based\\ Non-intrusive Load Monitoring Algorithm
}

\author{

	Elnaz Azizi, \emph{Student Member,~IEEE},
	Mohhamd~TH~Beheshti, \emph{Member,~IEEE}, \\ and Sadegh Bolouki, \emph{Member,~IEEE},
	
\thanks{E.~Azizi, M.~TH.~Beheshti and S.~Bolouki are with Department of Electrical and Computer Engineering, Tarbiat Modares University, Tehran, Iran. e-mails: \{e.azizi, mbehesht, bolouki\}@modares.ac.ir}

}

\begin{document}

\maketitle
\thispagestyle{empty}
\pagestyle{empty}

\begin{abstract}

	Non-intrusive load monitoring (NILM), aims to infer the power profiles of appliances from the aggregated power signal via purely analytical methods. Existing NILM methods are susceptible to various issues such as the noise and transient spikes of the power signal, overshoots at the mode transition times, close consumption values by different appliances, and unavailability of a large training dataset. This paper proposes a novel event-based NILM classification algorithm mitigating these issues. The proposed algorithm (i) filters power consumption signals and accurately detects all events, (ii) extracts specific features of appliances, such as operation modes and their respective power consumption intervals, from their power consumption signals in the training dataset, and (iii) labels with high accuracy each detected event of the aggregated signal with an appliance mode transition. The algorithm is validated using REDD with the results showing its effectiveness to accurately disaggregate low frequency measured data by existing smart meters.
\end{abstract}
\section{Introduction}

    Due to the unpredictable nature of both generation, caused by renewable energy resources, and consumer demand, maintaining the balance between generation and demand is one of the main challenges in smart grids \cite{8782552}. Residential demand-side management programs have thus emerged as a promising set of methods to strike such balance \cite{lu2019hybrid}. Among them, non-intrusive load monitoring (NILM), that is the process of extracting the power consumption profile or operating pattern of each appliance from the aggregated power consumption signal of a house using purely analytical methods, has gained a great deal of attention in recent years. Practical and efficient, NILM provides consumers with an opportunity to track the energy consumption of each appliance and voluntarily change their usage patterns to save energy and reduce the cost while maintaining their comfort which also results in higher stability and efficiency of the power grid \cite{ji2019non}.
	
	The concept of NILM was first introduced in 1992 by Hart \cite{hart1992nonintrusive}. Since then, a variety of analytical algorithms have been proposed to address the NILM problem. These algorithms employ various features and parameters such as voltage, current, and active and reactive power signals of a house. Since measuring the active power is cost-efficient, a majority of studies have focused on this feature alone \cite{rathore2018non}. NILM research based on the active power signal diverged into two main lines of study, (i) state-based algorithms that consider each appliance as a finite-state machine and disaggregate the total power signal based on the learned model of state transitions of appliances \cite{zhao2018improving} and (ii) event-based algorithms, which are based on the edges or considerable variations of the signal caused by turning ON/OFF of appliances or their other mode transitions \cite{lu2019hybrid}. Due to the low computational complexity of event-based techniques, they have proved more popular than the state-based ones \cite{liu2019low}.
	
    In designing an event-based algorithm, multiple challenges are involved. The first one lies in the event detection part caused by the presence of noise, spikes, uncertainties in the voltage of the grid, and overshoots in appliances' power consumption signals. The second challenge is closeness of different appliances' power consumption values which makes them somewhat indistinguishable. The third and last challenge is that high volume training datasets and ground-truth information about each appliance are scarce in practice, although a small amount of data can perhaps be collected for each residential building. Overcoming these challenges, this paper proposes an event-based NILM algorithm with competitive accuracy, which first via a novel method detects events in the power signals, then extracts specific features and information about appliances from their consumption profiles in a small training dataset, and finally utilizes them to disaggregate the aggregated power consumption signal.
	
\subsection{Related Work}

    The most well-known state-based NILM algorithms are the Hidden Markov Model (HMM) \cite{singh2017deep} and its variants such as Factorial HMM methods \cite{bonfigli2017non}. The main drawback of these methods is the requirement for a large training dataset to construct and learn the model. Computational complexities of these methods also increases exponentially by adding a new appliance \cite{he2019generic}. However, event-based NILM techniques which deal with the detected events of the aggregated signal and classify them have lower computational complexities in comparison with state-based ones \cite{azizi2020residential}.
    Recent research of event-based NILM falls into two main categories, namely unsupervised and supervised methods \cite{zhao2020non}. Unsupervised NILM algorithms, tackling the so-called blind source problem, deal with the case where no prior information about appliances is available. In these methods, events are detected and different clustering algorithms such as subtractive clustering \cite{henao2015approach} and $k$-means \cite{kong2016extensible} are applied to them. They detect different clusters of appliances without assigning a label to each cluster. Despite some success in the case where all appliances have only two (ON and OFF) modes, these algorithms have been ineffective in dealing with multi-mode appliances \cite{kong2016extensible, dinesh2019residential}.
    
    In contrast with unsupervised NILM methods, supervised algorithms such as NILM classification algorithms require prior metadata information about the number of appliances and their operation modes as well as a training dataset containing appliances' consumption profiles for a period of time. Considering modes of appliances as class labels, various classification methods such as KNN, multi-label classification \cite{tabatabaei2017toward,massidda2020non}, and deep learning \cite{singhal2018simultaneous} have been utilized in this field. These methods have proved to be significantly more accurate than their unsupervised counterparts, particularly in the presence of multi-state appliances \cite{singhal2018simultaneous}. However, their main drawback is the need for an enormous training dataset that is not in general feasible to collect \cite{yang2019systematic}. Therefore, extracting useful information from a small training dataset for the NILM classification problem has become a topic of great interest in the past few years \cite{dash2020appliance}.

\subsection{Contributions}
    
    
    This paper proposes a novel event-based NILM algorithm that minimizes the ground-truth data required, performs well in analyzing real data measured by existing meters, and remains efficient and accurate even for large numbers of appliances. Major contributions of this work are detailed below.
	
	\begin{itemize}[leftmargin=.165in]
	\item[{1)}] Event-based algorithms are highly dependent on detection of events. Therefore, the event detection algorithm used for the NILM purpose should be accurate in the sense that it should not miss any actual event or mistake fluctuations of the signal as an event. We propose in Section \ref{Pre-Processing} a novel Statistics-based method that filters the signal and detects events with a 100\% accuracy.
    \item[{2)}] For NILM as a classification problem, the number of operation modes of appliances and their respective power consumption values are key to assigning labels. In most of the existing literature, these modes are obtained by visually analyzing the appliances' power consumption signals in the training dataset. We introduce a clustering approach in Subsection~\ref{modeofapps}, using in part the linkage-Ward algorithm, which automatically extracts appliances' modes and their respective consumption values. Then, in Section~\ref{MLC}, a novel classification technique, with competitive accuracy, is established for the NILM problem employing previously extracted features.
    \item[{3)}] Existing classification algorithms consider appliances' power consumption values at each mode as their main characteristics. However, the appliance consumption pattern, its transitions between different modes, their ON duration period, and their probability of occurrence are also key information that can be used to distinguish two appliances with close power consumption values. Analyzing the training dataset in Section \ref{Information extraction}, we extract these features of appliances and utilized them for label refinement.


	\end{itemize}



\subsection{Paper Organization}

    The remainder of this paper is organized as follows. The terminology and problem statement are presented in Section \ref{S2}. In Section \ref{Pre-Processing}, the proposed signal filtering and event detection techniques are described. The feature extraction methods are then detailed in Section \ref{Information extraction}, followed by the proposed classification method in Section \ref{MLC}. The effectiveness and accuracy of the proposed algorithms are evaluated and compared with other algorithms using the REDD \cite{kolter2011redd} in Section \ref{Simulatin Study}. Finally, Section \ref{Conclusion} concludes the paper.
 
\section{Terminology and Problem Statement} \label{S2}

    In this section, we present the terminology and the event-based NILM classification problem considered in this paper. 
    
    
\subsection{Notions and Terminology}
    
     In this research, power refers to active power.
     {\it Operation modes} of an appliance refer to a fixed set of modes, including the OFF mode, in which the appliance can operate. Appliances are assumed to have two or more operation modes. The {\it operating mode} of an appliance is the mode in which the appliance is operating at a specific point in time. When no ambiguity results, the term {\it mode} is used to refer to an operation mode or operating mode. A {\it state} of an appliance is defined as its power consumption amount in one of its operation modes. Since this amount is assumed to vary at least slightly over time, a state is represented by a fixed closed interval within the set $\mathbb{R}$ of real numbers. One notices that there exists a state corresponding to each operation mode of an appliance.
    
    The {\it aggregated power consumption signal}, or simply the {\it aggregated signal}, refers to the sum of power consumption signals of all appliances of a house or specific appliances of interest. The term {\it non-intrusive load monitoring}, or NILM, in this work is then defined as extracting the sequence of operating modes of each appliance from the aggregated signal. This NILM problem is sometimes referred to as the NILM classification problem. While deducing individual power consumption signals of appliances from the aggregated signal is also a NILM problem, that one views as a NILM regression problem, it is not considered in this work.
    
    Given the power consumption signal of an appliance, an {\it event} is a change in the signal value caused by a mode transition of the appliance. Similarly, an event of the aggregated signal is a change in the signal value caused by a mode transition of any of the appliances contributing to the aggregated signal.

\subsection{Event-based NILM Problem}

    The event-based NILM (classification) problem can be described as the process of assigning proper labels to events of the aggregated power consumption signal, where the set of labels consists of all appliance mode transitions. It should be noted that a training set in the form of a set or sequence of events and their corresponding labels is in general not immediately available. Instead, it has to be derived from the given individual appliances' power consumption signals over a period of time. No additional information, such as the number of modes of each appliance or their nominal power consumption values, is available. It is assumed throughout the paper that the given signals are measured in discrete time.

\section{Signal Filtering and Event Detection}\label{Pre-Processing}

\begin{figure}[t]
        \centering
        \includegraphics[scale=0.18]{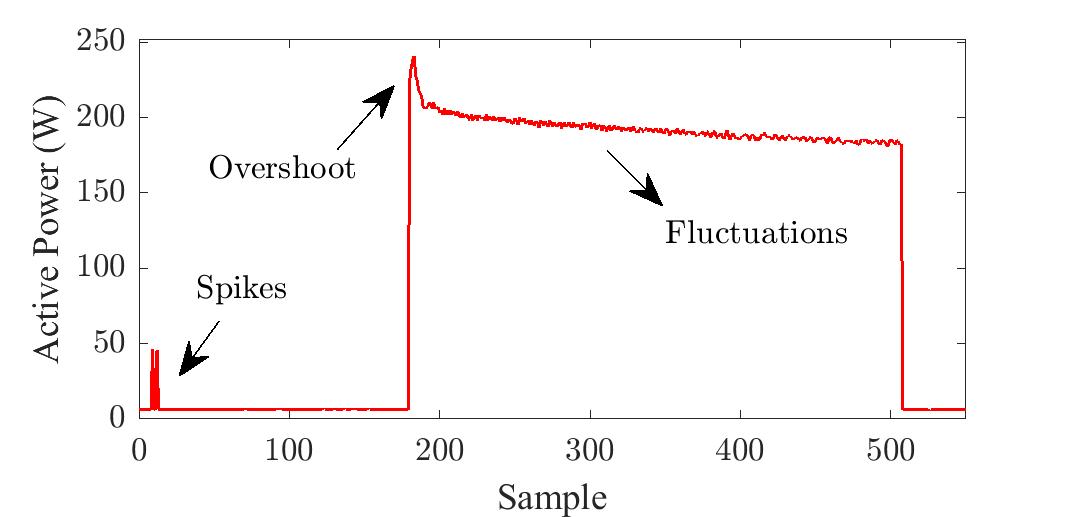}
        \caption{Typical outliers in consumption signal}
        \label{fig:overshoot_spikes_fluc}
    \end{figure}

    A fundamental part of event-based NILM is detecting events accurately. Event detection should be executed on individual appliances' power consumption signals in the training set as well as the aggregated signal in the test set. In  a  vast  majority  of  the  literature, an event is detected based on the difference between two consecutive sampled values. More precisely, if the absolute value of this difference is greater than a certain threshold, an event is considered to have occurred between the two sampling times. Existing threshold-based event detection techniques heavily rely on the threshold that is selected manually given the dataset in hand. Thus, they are not expected to perform as well on the meter's data of a different residential house. Beside this extensibility issue, there appears to exist a fundamental limit on the accuracy of threshold-based event detection techniques, which is caused by fluctuations of voltage in the power grid and noise, spikes, and the various ranges of overshoots in the signal, as shown in Fig.~\ref{fig:overshoot_spikes_fluc} \cite{henao2017approach, lu2017event}. In this section, we propose a novel Statistics-based algorithm that overcomes all these challenges and achieves 100\% accuracy in event detection.
    
    While the mainstream view of an event is a {\it significant} value change in the signal, our view of an event is an {\it uncommon} value change. Thus, considering a set formed based on value changes in the signal, we search for ``outliers'' of the set. As it will be explained later in this section, this set consists of the min/max ratios between consecutive sampled values of the signal, subtracted from 1. Of course, careful considerations should be made with regard to transient spikes and lengthy overshoots during mode transitions, as they would also be outliers of the formed set. Detecting these spikes and overshoots and filtering them will also prove significant for getting more accurate results in the event-based NILM problem.
    
    Our proposed event detection algorithm consists of three main steps, 1) outlier detection, 2) filtered signal construction and 3) event detection, which will be discussed in the following subsections.
\subsection{Outlier Detection}
    
    Different fields of research have been dealing with the outlier detection problem given a dataset and different methods have been proposed to address it \cite{kandanaarachchi2020normalization,chakhchoukh2016statistical}. We herein extend a Statistics-based outlier detection method to suit the NILM problem. As it is shown in Algorithm \ref{al1}, the algorithm starts with calculating the min/max ratio between any two consecutive sampled values of signal $P(t)$, subtracting them from 1, and saving them in a vector $M$. Then, the standard deviation of $M$ is computed. Finally, for any $t$, if $M(t)$ is grater than the calculated standard deviation, it is considered an outlier and $t$ is saved as the outlier occurrence instance in vector $M_o$ of outlier instances.
    \begin{algorithm} [t]
    \caption{Proposed algorithm for outlier detection.}\label{Times} 
    \textbf{Step 0:} Initialize the parameters, $i=1$, $M(:)=0$ and $M_o(:)=0$\\ 
    \textbf{Step 1:} Get signal $P(t)$, $t=1,\ldots,T$ \\ 
    \textbf{Step 2:} 
    \While{$t \leq T-1$}{
            $S(t)=\{P(t),P(t+1)\}$\\
            $M(t)=1-\frac{min(S(t))}{max(S(t))}$\\
            $t=t+1$}
    \textbf{Step 3:} Compute $sd$ as the standard deviation of $M$\\
    \textbf{Step 4:} Extract outliers' instances based on $sd$\\ 
    \While{$t \leq T-1$}{
        \If{$M(t)>sd$}{
        $M_o(i) = t$\;
        
        $i = i+1$\;}
        }
    \label{al1}
    \end{algorithm}
\subsection{Filtered Signal Construction}  \label{filtering}
\begin{figure}
        \centering
        \includegraphics[scale=0.58]{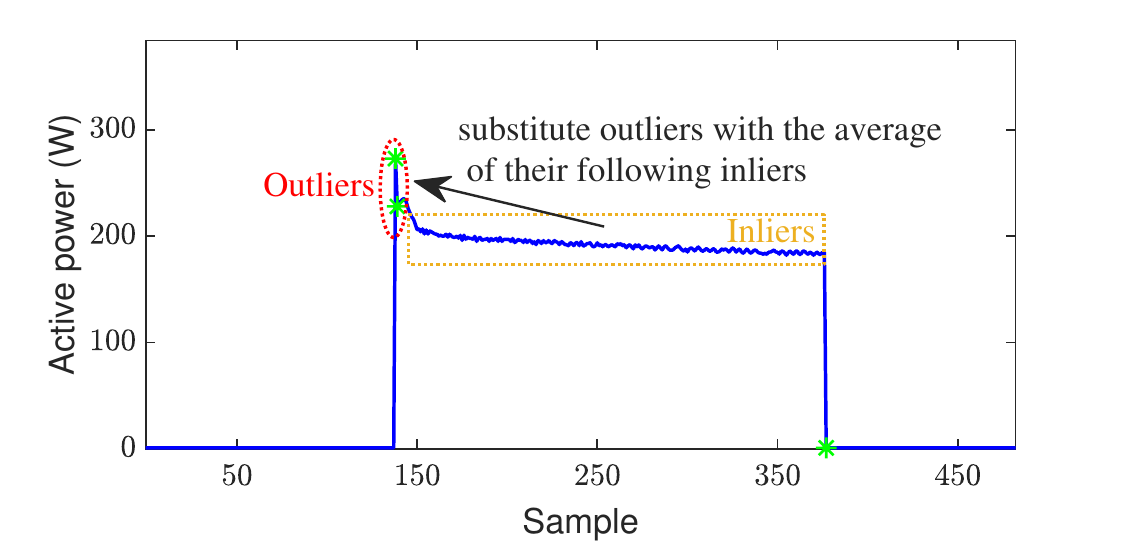}
        \caption{Construction of the filtered signal}
        \label{construction}
    \end{figure}

    Having performed outlier detection, outliers' instances are obtained, as well as instances that are not outliers, referred to as {\it inlier} instances. In this so-called filtering step, the aim is to flatten spikes and overshoots of the signal. To achieve this aim, the signal value at each outlier instance is substituted with the mean of the signal values at the following consecutive inlier instances, as shown in Fig.~\ref{construction}. You may note that, unlike for spikes and overshoots, the signal values at actual event times will not experience significant change in the filtering step.


     %
%
 %

\subsection{Event Detection} 

    In the final step, to detect events, Algorithm \ref{al1} is applied to the filtered signal. Now, all detected outlier instances are considered as event instances. We note that transient spikes and overshoots of the original signal have already been flattened in constructing the filtered signal, meaning that they can no longer be mistaken for events.

\section{Feature Extraction} \label{Information extraction}

    The most common specific feature of appliances utilized in NILM algorithms is their power consumption values at their operation modes. However, since different appliances may have operation modes with close power consumption values, an effective load disaggregation algorithm should use additional appliance features broadly called {\it finger prints} in \cite{ma2017toward}. In this section, we propose different methods to extract operation modes of appliances and additional features from a small training dataset.

\subsection{Modes and States of Appliances} \label{modeofapps}
    
    States of an appliances, defined as the power consumption interval corresponding to its operation modes, are its most useful features widely used for the NILM purpose. Therefore, detecting the number of modes of an appliance and their corresponding states plays a crucial role in the accuracy of NILM. As opposed to most of the existing literature that extracts an appliance's modes/states visually using its power consumption values in the training set or by using the datasheet of the appliance, we propose a novel clustering-based approach for appliance modes/states extraction from its power consumption signal in a systematic fashion.
    
    Our approach is based on the linkage-Ward (LW) clustering algorithm \cite{alpaydin2020introduction}. The objective function of the LW algorithm is the squared sum of distances between data points and the their cluster centroids. 
    The LW clustering algorithm first treats 
    each data point as a cluster of its own, which means that the initial value of the objective function is 0. Then, clusters are merged together, one pair at every stage, based on the following merging policy: clusters $A$ and $B$ are merged if $\Delta(A,B)$ is a minimum among all pairs of clusters,
    \begin{equation}\label{linkage}
    \begin{array}{cc}
    \Delta (A,B)
    &\hspace{-.85in}= \underset{\tiny p \in A\cup B}{\mathlarger{\mathlarger{\sum}}}\| p - m_{A\cup B}\|^2\vspace{.1in}\\
    &\hspace{-.1in}-\underset{\tiny p\in A}{\mathlarger{\mathlarger{\sum}}}\,\| p - m_{A}\|^2 - \underset{\tiny p \in B}{\mathlarger{\mathlarger{\sum}}}\,\| p - m_{B} \|^2\vspace{.1in}\\
    &\hspace{-.8in}=\frac{n_{A}n_{B}}{n_{A}+n_{B}}\left\| m_{A}-m_{B} \right\|^2,
    \end{array}
    \end{equation}
    where $m_{A}$, $m_{B}$, and $m_{A\cup B}$ are centroids of clusters $A$, $B$ and ${A\cup B}$ respectively; and $n_A$ and $n_B$ are the size of clusters $A$ and $B$. One notices that $\Delta (A,B)$, which is non-negative, is in essence the cost of merging clusters $A$ and $B$. Thus, the value of the objective function at any given stage is the total cost of all the merging up to that stage. Analyzing the increasing objective function, the {\it elbow method} \cite{azizi2020residential} is often used to determine the optimal number of clusters, that is to determine when to stop merging. Roughly speaking, according to the elbow method, merging stops when it becomes too costly compared to the merging at the previous stage.
    
    It can be seen from the definition of $\Delta(A,B)$ in \eqref{linkage} that the combination of the LW algorithm and the elbow method is susceptible to unbalanced data. More precisely, when the LW algorithm is close to reaching the optimal number of clusters, the elbow method discourages stopping where two or more small clusters exist since merging them is now not relatively ``too costly'' even though their centroids may be far apart. For the mode/state extraction purpose, this could be a significant issue, as infrequently occurring modes of an appliance may be lumped into one mode with a wide-ranging power consumption, which would undermine any attempt of NILM. Therefore, one should take advantage of the LW algorithm in such a way to suit the NILM purpose by discouraging merging of clusters that are far apart. Thus, the following algorithm is suggested for mode extraction.
    
\begin{figure*}
        \centering
        \includegraphics[scale=0.66]{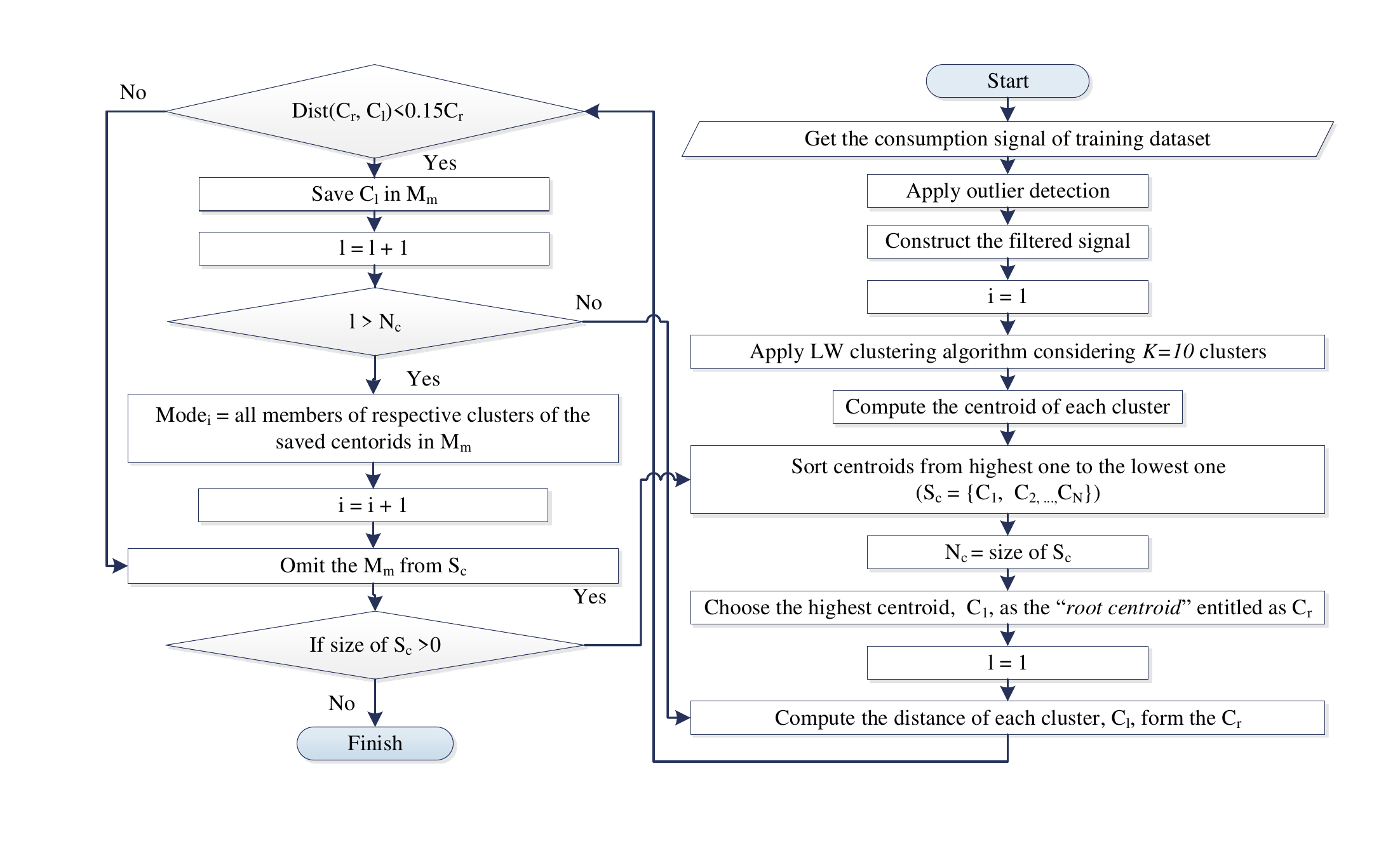}
        \caption{Flowchart for the proposed mode extraction algorithm}
        \label{Mode_extraction}
    \end{figure*}

    After filtering the power consumption signal of each appliance in the training dataset, the LW algorithm is applied to the signal's data considering $K$ clusters, where $K \geq 10$. The cluster centroids are then computed 
    and sorted in descending order. From this stage forward, a different merging policy,  entitled as {\it distance-based} policy, is adopted that only depends on the distance between cluster centroids. It starts by considering the cluster with the highest centroid as the {\it root cluster}. If its centroid's distance to the next highest centroid is less than 15\% of the root cluster's centroid, the two clusters merge and the step is repeated considering the merged cluster as the new root cluster. Otherwise, the cluster with the next highest centroid to the root cluster's centroid is considered as the new root cluster and the step is repeated. The algorithm terminates when no further merging can take place. The flowchart of the proposed mode extraction method is illustrated in Fig.~\ref{Mode_extraction}, where centroids of the root cluster and the cluster with the next highest centroid are denoted as $C_{r}$ and $C_{j}$.
\subsection{Transition Intervals of Appliances}
    
    Consider an arbitrary mode transition of an appliance from a (relatively) high state $H=[H^{min},H^{max}]$ to a (relatively) low state $L=[L^{min},L^{max}]$. Then, the interval of this transition is defined as
    \begin{equation}\label{intervals_trans}
	    [H^{min}-L^{max},H^{max}-L^{min}].
	\end{equation}

\subsection{Transition Participation Indices}\label{Appliance_Par}

    Two appliances with some overlapping transition intervals can be difficult to separate in the aggregated signal. The {\it participation index} of their transitions defined as \eqref{p_co}, obtained from each appliance's usage pattern in the training set, may help separate these appliances. Given a transition $T$ of an appliance, its participation index $P_{par}$ is defined as
    \begin{equation}\label{p_co}
        P_{par} = \frac{1}{N_{days}} \mathlarger{\mathlarger{\mathlarger{\sum}}}_{day=1}^{N_{days}}\frac{\text{number of }T \text{ in } day}{\text{number of events in }day}.
    \end{equation}
    where $N_{days}$ is the total number of days of training dataset in which the transition $T$ happened. In other words, this parameter measures the daily average of contribution of a specific transition of an appliance in events of the total signal.

\subsection{Additional Features of Appliances} \label{Features}
    
    Analyzing the power consumption signals of appliances in the training dataset, one observes that some of the appliances exhibit very specific behavioral patterns. Some of these features, that can further help improve the accuracy of load disaggregation, are listed below.
    
\begin{figure}
        \centering
        \includegraphics[scale=0.4]{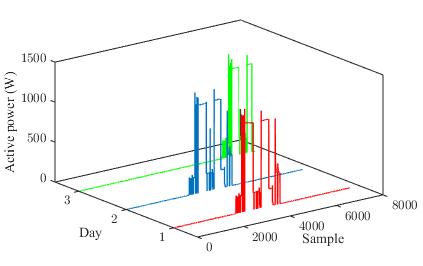}
        \caption{Daily usage pattern of the dishwasher}
    \label{DW_3D}
    \end{figure}

    \begin{itemize}[leftmargin=*]
        \item As shown in Fig.~\ref{DW_3D}, the dishwasher exhibits the same pattern when it is ON. As this pattern is complex, using it may hurt efficiency of NILM. However, one notices that in any given day, the dishwasher operates in either all or non of its modes. This simple characteristic will prove significant for NILM.
        \item One observes that not all possible mode transitions of multi-mode appliances can ever occur.
        \item Some appliances appear to have unique mode transition overshoots in their signals.
        \item The time period between two consecutive ON samples (with OFF samples in between) of different appliances are significantly different.
    \end{itemize}
\section{Classification} \label{MLC}

    In this section, a novel classification algorithm is proposed to address the NILM problem, that is to determine each event in the aggregated power consumption signal corresponds to which appliance mode transition. To label a given event, the classifier first obtains all transition intervals containing the value of that event. As a simple example, consider two ON/OFF appliances 1 and 2, whose ON modes correspond to states [890,1000] and [970,1050], respectively. Given an event with value 980 in the aggregated test signal, the classifier first associates this event with OFF-to-ON transition of appliance 1 as well as that of appliance 2. Afterwards, taking other specific features of each appliance into considering along with the test signal's behavior near the event, a single label is assigned to the event.
    
    Defining $N_e$ and $N_T$ as the total number of events in the test signal and the number of all possible appliance mode transitions, respectively, let $L_e$ be an ${N_T\times N_e}$ binary-valued matrix, with columns corresponding to events and rows corresponding to mode transitions, which represents predicted labels for events of the test signal. Obviously, an element 1 of $L_e$ indicates that its row's corresponding transition is the predicted label of its column's event. The proposed classification algorithm is detailed in 4 steps below.
    
    \noindent \textbf{Step 1:} Given an event of the test signal and a mode transition of an appliance, if the event value is within the transition interval, the respective element of $L_e$ is labeled 1. Otherwise, it is labeled 0. It should be clear that since transition intervals may overlap, some events may be assigned multiple labels in this step. We also point out that 
    some events may remain unlabeled, which means that matrix $L_e$ may have some all-zero columns. Each of these unlabeled events is then labeled the transition whose interval is closest to the event value. We note that the distance between a value and an interval is calculated as the minimum distance between the value and any point within the interval.

\begin{figure}\label{DW}
        \centering
        \includegraphics[scale=0.38]{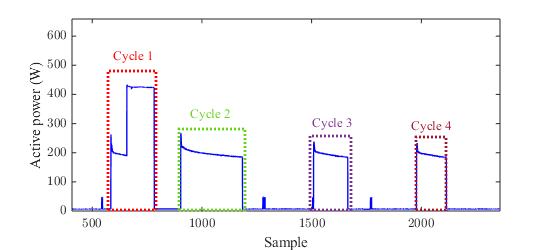}        \caption{Daily usage pattern of the refrigerator vs its cycles}
        \label{Daily}
    \end{figure}

    \noindent \textbf{Step 2:} Analyzing the daily aggregated signal, it can be observed that in a majority of time samples, appliances are in their OFF modes. With that in mind, one should focus on parts of the aggregated signal where at least one appliance is ON. In particular, given the aggregated signal, one obtains its {\it cycles}, where each cycle starts with an event succeeding an all-OFF sample and ends with the nearest event preceding an all-OFF sample. Fig. \ref{Daily} illustrates how cycles of an aggregated signal are derived. Over each cycle, the labels assigned to the events should be {\it compatible}.
    
    To clarify what is meant by labels compatibility over a cycle, one thinks of an undirected graph in which nodes represent all mode vectors $\theta$, where each element $\theta_a$ of $\theta$ is a mode of appliance $a$. Two nodes are then connected by an edge if their corresponding vectors differ in exactly one element. It should be clear that an edge represents a single appliance mode transition, while noting that two different edges may correspond to the same transition. Now, a sequence of labels over a cycle are said to be {\it compatible} if starting from the all-OFF node of the graph, one can walk according to the labels in the sequence and terminate at the all-OFF node. In other words, labels/transitions over a cycle are compatible if they form a cycle in the graph constructed above.
    
    Using the compatibility condition described, the multiple labels assigned to some events can be narrowed down as some labels are deemed inadmissible. More precisely, a label within a cycle is removed if it is not part of any compatible sequence of labels over the cycle. As an example, in cycle~1 of Fig.~\ref{Daily}, after Step~1, assume that the first and second events are assigned single labels, that are mode transitions of appliance 1; while the third event is assigned two labels, one a mode transition of appliance 1 and one a mode transition of appliance 2. Then, in Step~2, the mode transition of appliance 2 is ruled out as a label of the third event since it is not part of any compatible sequence of labels from Step~1 over cycle~1. In other words, appliance 2 cannot possibly be ON when the third event occurs.
    
    \noindent \textbf{Step 3:} After Step~2's cycle-based label refinement, each event may still be assigned multiple labels. Considering extracted specific features of some appliances as described in \ref{Features}, some labels are removed from the multi-labeled events.

    \noindent \textbf{Step 4:} Finally, based on participation indices of appliance transitions explained in \ref{Appliance_Par}, the most probable label is chosen for the multi-labeled events.

\section{Simulation Study} \label{Simulatin Study}

\begin{figure}[t]
	        \centering
	        \includegraphics[scale=0.4]{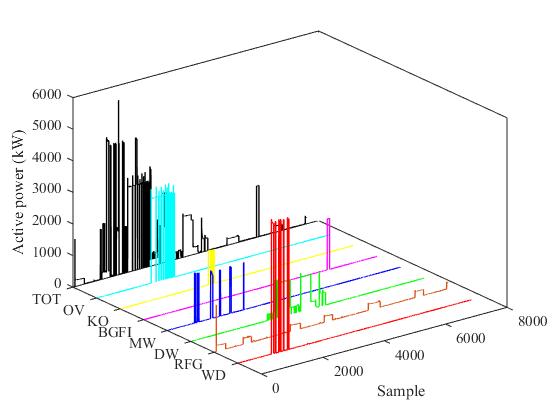}
	        \caption{Total consumption vs the consumption of each appliance}
	        \label{3D_test_days}
	    \end{figure}

    In this section, the accuracy and effectiveness of our algorithms proposed in Sections \ref{Pre-Processing}--\ref{MLC} for the NILM purpose are evaluated by applying them to a low-frequency dataset, gathering which is practical as it can be done using existing smart meters \cite{El2020Residential}. The dataset considered consists of 28 days of power consumption data for seven appliances of house 1 in the REDD \cite{kolter2011redd}. These appliances are listed as oven (OV), microwave (MW), kitchen outlets (KO), bathroom GFI (BGFI), washer/drier (W/D) with a high consumption state, refrigerator (RFG), and dishwasher (DW). The first five listed appliance only have ON and OFF modes, while the last two have more than two operation modes. Fig.~\ref{3D_test_days} shows the power consumption signals of all seven appliances in a day of the dataset.
    To measure the accuracy of the proposed classification, the $F_{measure}$ is used \cite{xu2019new},
	    \begin{equation}
        F_{measure} = \frac{2\times TRP \times RC}{TRP+RC},
	    \end{equation}
        \begin{equation}
        TPR = \frac{TP}{TP+FN},~RC = \frac{FP}{FP+TN},
	    \end{equation}
    \noindent where $TRP$ and $RC$ show the precision and recall; $TP$ is true positive, $FP$ is false positive, $TN$ is true negative, and $FN$ is false negative.
    
    In the following subsections, first the proposed filtering end event detection method are applied to training and test dataset. Then, based on the filtered signal of each appliance, their specific features are extracted. Finally, considering these features the proposed classification technique is utilized to disaggregate the test signal.
    
\subsection{Signal Filtering and Event Detection}
    
\begin{figure}[t!]
	    \centering
	    \includegraphics[scale=0.62]{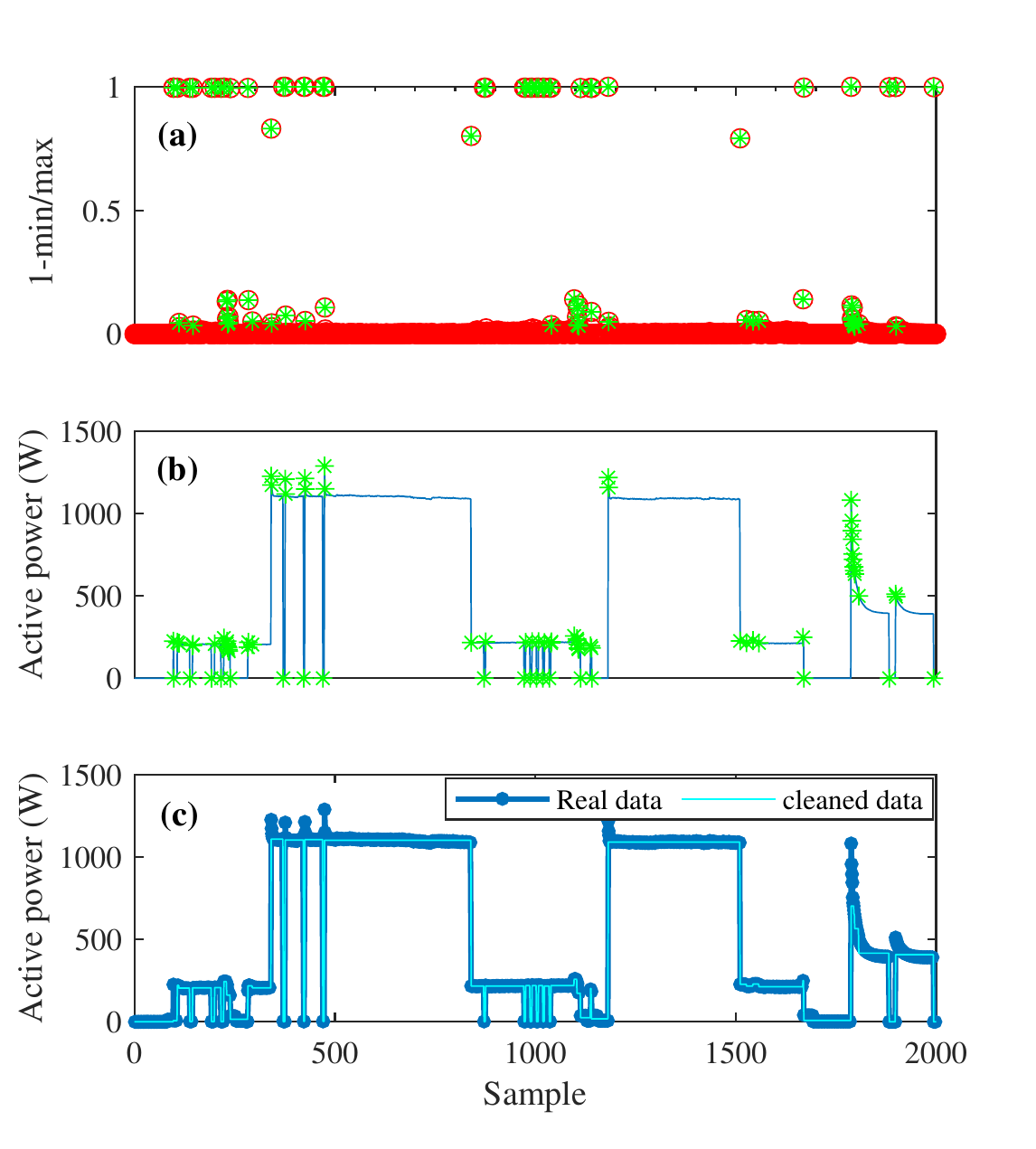}
	    \caption{Filtering the dishwasher's signal: (a) inliers vs. outliers in 1-min/max ratios of consecutive samples, (b) detected outliers in the power consumption signal, (c) the orignal signal vs. the filtered signal}
	    \label{fig:outliers}
	\end{figure}

    The filtering  and event-detection method of Section \ref{Pre-Processing} is applied to individual appliances' power consumption signals in the training dataset as well as the aggregated signal in the test dataset. As an example, Fig.~\ref{fig:outliers} illustrates the outliers of the signal, the overshoots and spikes in the signal and the constructed filtered signal of the dishwasher for a period of time, respectively.

\subsection{Feature Extraction}

    Having individual appliances' signals filtered, their features are extracted via methods of Section \ref{Information extraction} as discussed below.
    
\subsubsection{States of Each Appliance}

 
    Applying the LW-based clustering method of Subsection \ref{modeofapps} to the filtered signal of each appliance, its modes and their corresponding states are obtained. One recalls that the merging policy is that of the LW algorithm until 10 clusters are obtained, then changes to that of distance-based policy. 
    As an example, Table~\ref{DW_10_clusters} shows the 10 obtained clusters for modes/states of the dishwasher before the merging policy changes from the LW method, and Table~\ref{statesTable} shows the final obtained states of all appliances.

    \begin{table}\centering 
    \caption{Minimum, maximum, and centroid of the 10 LW-obtained clusters for dishwasher's modes/states.}
    \begin{tabular}{|c|c|c|c|}
    \hline
                                       & min  & max  & centroid \\ \hline
    \multirow{10}{*}{Active power (W)} & 154  & 183  & 168      \\ \cline{2-4} 
                                       & 198  & 208  & 205      \\ \cline{2-4} 
                                       & 210  & 231  & 219      \\ \cline{2-4} 
                                       & 236  & 261  & 236      \\ \cline{2-4} 
                                       & 398  & 420  & 449      \\ \cline{2-4} 
                                       & 416  & 496  & 489      \\ \cline{2-4} 
                                       & 500  & 598  & 589      \\ \cline{2-4} 
                                       & 643  & 737  & 680      \\ \cline{2-4} 
                                       & 1078 & 1110 & 1099     \\ \cline{2-4} 
                                       & 1115 & 1247 & 1173     \\ \hline
    \end{tabular}
    \label{DW_10_clusters}
    \end{table}	
    \begin{table}\centering
    \caption{States corresponding to non-OFF modes of appliances}
    \begin{tabular}{|c|c|}
    \hline
    App       & States    \\ \hline
    DW      & 157-183,198-261,398-496,643-737,1078-1247 \\ \hline
    RFG    & 153-183,185-260,415-425\\ \hline
    MW      & 1439-1598        \\ \hline
    BGFI    & 1580-1620         \\ \hline
    KO & 1064-1087 \\ \hline
    W/D    & 2641-3000 \\ \hline
    OV            & 4138-4157  \\ \hline
    \end{tabular} 
    \label{statesTable}
    \end{table}

\subsubsection{Transition Intervals of Appliances}    

    Intervals of probable mode transitions of all appliances are now derived using \eqref{intervals_trans}. When a mode transition is from or to the OFF mode of an appliance, it simply coincides with the state corresponding to the other appliance mode involved in that transition. Considering probable transitions of multi-mode appliances, non-trivial transition intervals [817-1049] and [155-240] should be considered for DW and RFG, respectively.
    
\subsubsection{Transition Participation Index}
    
    One recalls that transition participation index is only used to separate appliance transition with overlapping intervals. Table \ref{P index} shows participation index for different groups of the overlapped appliances.
\begin{table}[]\centering
\caption{Transition participation index}
\begin{tabular}{|c|c|c|c|}
\hline
                                                                                            & Group              & Corresponding app     &  $P_{par}$\\ \hline
\multirow{4}{*}{\begin{tabular}[c]{@{}c@{}}Overlapping\\ transitions \\ group\end{tabular}} & \multirow{2}{*}{1} & BGFI &  0.11\\ \cline{3-4} 
                                                                                            &                    & MW   & 0.20 \\ \cline{2-4} 
                                                                                            & \multirow{2}{*}{2} & KO   & 0.17 \\ \cline{3-4} 
                                                                                            &                    & DW   &  0.73\\ \hline
\end{tabular} \label{P index}
\end{table}
    
    \begin{table}[t!]\centering
    \caption{$F_{measure}$ for different methods}
    \begin{tabular}{|c|c|c|c|}
    \hline
     App & \cite{dash2020appliance} & ML-KNN \cite{li2018residential} & Proposed method  \\ \hline
    DW       &  0.59 &  -  & 0.96  \\ \hline
    RFG      & 1 & 0.88 & 0.81  \\ \hline
    MW      & 0.40 & 0.76 & 0.82 \\ \hline
    BGFI    & 0.97 & 0.88 & 0.91\\ \hline
    KO      & 1 & 0.84 & 0.86 \\ \hline
    WD     & 0.92 & 0.94 &1 \\ \hline
    OV      & - & 0.86 & 1   \\ \hline
    LIGHT    & - & 0.76 & -\\ \hline
    AVERAGE  & 0.81 & 0.80 & 0.90 \\ \hline
    \end{tabular} \label{TP_FP_before_post}
    \end{table}
\vspace{-.005in}    
    
\subsection{Classification}
        
    To apply the proposed classification method, events of the filtered test signal are detected. Then, the classification-based algorithm is applied to detected events based on the transition intervals of appliances. Finally, after cycle-based label refinement, to refine the remained multiple labeled events and choose the most probable label for each event, the following specific features of appliances are considered:
    \begin{itemize}[leftmargin=.165in]
        \item[{1)}] Based on the power consumption signal of the dishwasher, one observes that during every ON cycle, all its modes occur. Thus, if the transition with interval [643-737] of the dishwasher, which happens to not overlap with any other transition interval, is not detected, the dishwasher can be assumed OFF during that whole day. Consequently, all candidate labels/transitions involving a non-OFF mode of the dishwasher in that day can be ruled out.
        \item[{2)}] Three modes for the refrigerator have been detected, namely the OFF mode, the lower ON mode, and the higher ON mode. It is observed in the training dataset that the transition from the OFF mode to the higher ON mode never happens. Thus, all candidate labels of such can be removed.
        \item[{3)}] Overshoot values appearing in the refrigerator's signal are higher than 500~W as opposed to those of the dishwasher. This can be used to separate transitions of the refrigerator and dishwasher with overlapping intervals.
    \end{itemize}
    Applying aforementioned features, for each remaining event with multiple labels, the participation index is calculated for overlapped appliances separately. The appliance which has close $P_{par}$ to the calculated participation index of appliances in training dataset, is assigned to events.
    Table~\ref{TP_FP_before_post} shows the high accuracy of our classification method for each appliance in comparison with the results of \cite{dash2020appliance} and \cite{li2018residential}. Keeping in mind that a higher number of appliances should diminishes the accuracy of NILM, one notes that the number of appliances considered in this work and \cite{dash2020appliance} are seven and six. On the other hand, considering multi-mode appliances such as dishwasher increases the complexity of disaggregation. However, the accuracy of our proposed method in which we have considered dishwasher is higher than \cite{li2018residential} which did not considered dishwasher in appliances' set.

\section{Concluding Remarks and Future work} \label{Conclusion}
    
    In this paper, we have proposed a novel classification method to address the NILM problem given a small dataset. The proposed algorithm has three main phases: 1) filtering training and test signals and accurately detect their events using a statistics-based method, 2) extracting features of appliances, most notably their modes and states via a clustering approach that in part uses the LW clustering method, 3) proposing a classification algorithm labeling events of the aggregated test signal with mode transitions of appliances, where various features and techniques are utilized to enhance its accuracy.
    The proposed event detection and modes/states extraction methods have been done in a systematic fashion in such a way to perform well for any set of power consumption data. The proposed filtering method, feature extraction techniques, and event-based NILM classification algorithm have been validated using the REDD. Results show 100\% accuracy in event detection. Juxtaposing the results of our classification algorithm with two recently introduced event-based NILM methods indicate a relatively high accuracy of our algorithm.
    
    Reconstructing the power consumption signals of appliances, which can be cast as a regression problem as opposed to the classification problem we considered in this work, is one of the main challenges of event-based NILM problems. We aim to modify our algorithms, mainly the one in Section \ref{MLC}, to address the reconstruction problem. Moreover, due to the lack of a training dataset for each residential building, we wish to move a step further to use transfer learning to circumvent the training phase of the proposed method, with the practical assumption that nominal values for appliances' power consumption are given.

\balance
\bibliography{NILM_Second_paper}
\bibliographystyle{IEEEtran}

\end{document}